\documentclass[12pt]{iopart}
\usepackage{epsfig,harvard,amsbsy,bm}
\newcommand{\isum}%
{\mathop{\hbox{$\displaystyle\sum\kern-13.2pt\int\kern1.5pt$}}}
  \newcommand{\la}{\langle}
  \newcommand{\ra}{\rangle}
\newcommand{\ba}{\begin{eqnarray}}
\newcommand{\ea}{\end{eqnarray}}
\renewcommand{\br}{\begin{eqnarray*}}
\newcommand{\er}{\end{eqnarray*}}
\newcommand{\be}{\begin{equation}}
\newcommand{\ee}{\end{equation}}
  
\renewcommand{\k}{{\bm k}}
\renewcommand{\b}{{\bm b}}
  \newcommand{\p}{{\bm p}}
\renewcommand{\e}{{\bm e}}
\renewcommand{\d}{{\bm d}}
  \newcommand{\n}{{\bm n}}
  
\renewcommand{\r}{{\bm r}}

\newcommand{\hs}{\hspace*}

\begin{document}
\bibliographystyle{jphysicsB}

\title[Two-photon double ionization of helium]
{Convergent close-coupling calculations of  two-photon double 
ionization of helium.}

\author{A. S. Kheifets
\footnote[1]{Corresponding author: A.Kheifets@.anu.edu.au}
and 
I. A. Ivanov\footnote[2]{On leave from the Institute of
Spectroscopy, Russian Academy of Sciences}
}
\address{Research School of Physical Sciences and Engineering,
The Australian National University,
Canberra ACT 0200, Australia}

\date{\today}
\begin{abstract}
We apply the convergent close-coupling (CCC) formalism to the problem
of two-photon double ionization of helium. The electron-photon
interaction is treated perturbatively whereas the electron-electron
interaction is included in full.  The integrated two-photon double
ionization cross-section is substantially below non-perturbative
literature results. However, the pattern of the angular correlation in the
two-electron continuum is remarkably close to the non-perturbative
time-dependent close-coupling calculation of Hu {\em et al}
[J. Phys. {\bf B38}, L35 (2005)]
\end{abstract}
\pacs{42.50.Hz 32.80.-t 32.80Fb }
\maketitle

\section{Introduction}
\label{Sec1}

One-photon double ionization of He has been a subject of intense
theoretical and experimental studies over the past decade.  As a
result, a complete understanding of this reaction has emerged both in
terms of underlying physical processes and the precise knowledge of
magnitudes and shapes of the cross-sections over a wide range of
photon energies
\cite{BS00,KA00,AH05}. 

Two-photon double ionization (TPDI) of He represents a new level of
complexity and brings new challenges, both to theory and experiment.
Extremely intense  VUV radiation is needed to study TPDI of He.
Only very recently, such a radiation has become available from the
free-electron laser sources \cite{LdeCG05} and high harmonics generation
\cite{NHT05}. Dynamics of the TPDI is richer as compared to
one-photon double ionization because of interplay of the $S$ and $D$
continua. Also, the target description becomes more involving since
the knowledge of the intermediate state as well as the initial and the
final states is needed. More importantly, the target states can be
modified by the strong laser field.

There have been several reported calculations of the total TPDI
cross-section of He at various photon energies
\cite{PR98,NL01,PMM01,MHN01,CP02a,FdH03,PBLB03,HCC05}. 
Although numerical values of the cross-sections vary depending on the
theoretical model and assumed characteristics of the laser field,
there is  some consensus between several calculations (see
\citeasnoun{HCC05} for detail). It has also been well
established that a non-perturbative treatment of the electron-photon
interaction is needed to obtain accurate total cross-sections.

As compared to the total cross-section, the present knowledge of the
differential cross-sections of the TPDI of He is limited. Only
time-dependent close-coupling (TDCC) calculations have been reported
at photon energies close to double ionization threshold
\cite{CP02a,HCC05}. In the meantime, the fully-resolved
triple-differential cross-section (TDCS) contains the most detailed
information on the two-electron break-up process. In the case of
one-photon double ionization, the analysis of the TDCS brought a wealth
of information allowing to clearly separate various mechanisms of photo
double ionization \cite{BS00,KKB02}

So there is a compelling motivation to perform further studies of He
TPDI TDCS at a wide photon energy range. To serve this purpose, we
apply the convergent close-coupling (CCC) formalism to describe the
two-electron continuum following the TPDI of He. This method has been
extensively used to study one-photon double ionization
\cite{KB96,KB98b,KB98e}. In the two-photon case, a further development 
of the method is needed. Firstly, integration over all the
intermediate states of the target is required following absorption of a
single photon. This involves evaluation of the ill-defined
continuum-continuum dipole matrix elements. To circumvent this
difficulty, we perform our calculations in the Kramers-Henneberger
gauge of the electromagnetic field \cite{IK05b}.  As an alternative
and much less time consuming method, we use the closure approximation
to carry out summation over all the intermediate target states. As the
result of this procedure, we end up with evaluation of the monopole
and quadrupole matrix elements between the correlated ground state and
the CCC final state. This procedure has been worked out in an earlier
paper on the second Born treatment of the electron impact
ionization-excitation and double ionization of He \cite{K04}.

Secondly, as compared to the weak-field single-photon double
ionization, the theoretical description of TPDI requires an account of
the field modification of the target states. Conceptually, this
procedure has been developed in a way which can incorporate the CCC
formalism \cite{IK05}. However, at the present stage, we are unable to
perform fully converged numerically accurate calculations of this type
due to limitations of computer power. Leaving this task for the
future, we report in the present paper the CCC calculations of the He
TPDI TDCS in the perturbative regime when interaction of the target
and the electromagnetic field is treated to the second order of the
perturbation theory. Although this type of calculation would be
clearly inadequate to obtain accurate total TPDI cross-sections, it
can be useful in evaluating the angular correlation pattern in the
two-electron continuum. In this respect, the present model is somewhat
analogous to the asymptotic Coulomb treatment of the one-photon double
ionization \cite{SKS94,MB93}. The asymptotic three-body Coulomb (3C)
wave function is only correct at large distances from the nucleus. The
fact that calculations with this wave function reproduce correctly the
angular correlation pattern in the two-electron continuum tells us
that it is the large distances which are responsible for forming this
pattern. In the meantime, the magnitude of the double photoionization
cross-sections calculated with the 3C wave function is typically off
by a significant factor which means that the photoelectron flux
originates from vicinity of the nucleus where the 3C wave function is
incorrect. The same benefits and disadvantages of such an approximate
model can be seen in the present work. Indeed, the fact that the
angular correlation pattern of the two-photon double ionization can be
calculated correctly within a lowest order of the perturbation theory
tells us that it is the inter-electron correlation in the final state
that is responsible for forming this pattern. In the meantime, the
total cross-section which is incorrect in such a perturbative
calculation is sensitive to the detail of the electromagnetic
interaction and requires a fully non-perturbative treatment.

The paper is organized as follows. In Sec.~\ref{Sec2} we give an
outline of the formalism. In Sec.~\ref{Sec3} we give our numerical
results which are obtained both with the CCC integration over the
intermediate target states and the closure approximation. Subsections
\ref{Sec3a}, \ref{Sec3b}, \ref{Sec3c} contain the analysis of the
integrated and differential cross-sections and the symmetrized
ionization amplitudes, respectively. We conclude in Sec.~\ref{Sec4} by 
outlining our future directions in this project.

\section{Formalism}
\label{Sec2}

\subsection{Second order perturbation theory}

We use the following second order perturbation theory expression for
the TPDI TDCS:
\be
\label{TDCS}
\hs{-2cm}
\frac{d^3 \sigma_{2\gamma}}{d\Omega_1 d\Omega_2\,dE_2}=
C_{2\gamma}
\omega^\beta  \
\Bigg|
\sum_i
\int \! d^3p 
\frac
{\langle 
\Psi_f(\k_1) | \ \e\cdot\d \ |\Psi_i(\p)\rangle
\langle \Psi_i(\p) |\ \e\cdot\d \  |\Psi_0\rangle}
{E_0+\omega-p^2/2-\varepsilon_i+i\delta}\\
\langle \k_2| f\rangle
\Bigg|^2
\nonumber
\ee
Here the two-photon ionization constant
$C_{2\gamma}=8\pi^3 c^{-2}a_0^4\tau=2.505\times10^{-52}$ 
cm$^2$ s$^{-1}$
\cite{TB93}. 
The vector $\e$ represents a linearly polarized light.  Generalization
for an arbitrary polarization is straightforward \cite{MMM96}. The
dipole operator $\d = \d_1+\d_2$ where $\d_\alpha=\r_\alpha$, ${\bm
\nabla}_\alpha$ and $Z\r_\alpha/r_\alpha^3$ in the length, velocity and
acceleration gauges, respectively, the nucleus charge $Z=2$ for
helium. The exponent $\beta$ depends on the gauge of the
electromagnetic interaction, $\beta=2$ in the length gauge.

In the CCC formalism, we represent the two-electron state by a
close-coupling expansion over the  channel states each of
which is composed of a target pseudo state $f$ and a Coulomb wave
$\k$:
\be
\label{E7}
\Psi_f(\k)
=
| \k f \ra
+\sum_j\isum d^3k'
 {\la \k f|T|j \k' \ra \over E-k'^2/2-\varepsilon_j+i0} \ 
| \k'j \ra
\ .
\ee
Here $\la \k f|T|j \k' \ra$ is half-on-shell $T$-matrix which is found
by solving a set of coupled Lippmann-Schwinger equations \cite{BS95b}.
The final state with two electrons in the continuum in \Eref{TDCS} is
obtained by projecting the positive energy pseudostate of the matching
energy $\varepsilon_f=k_2^2/2$ on the Coulomb wave $|\k_2\ra$.

Using the following partial wave expansions for the $T$-matrix
$$
\la \k f|\,T\,|j\k'\ra=
\sum\limits_{{L,L',J\atop M,M',M_J}}
C_{LM,\,l_fm_f}^{JM_J}
C_{L'M',\, l_jm_j}^{JM_J}
Y_{LM}(\n)Y_{L'M'}^*(\n') \ \
\la kL \ n_fl_f \| T_{J} \|n_jl_j \ k'L'\ra
$$
and the dipole matrix element:
$$
\la \k f|\ \e\cdot\d \ |\Psi_0\ra
=
\sum_{M_P} e_{M_P} \ 
\sum_{lm}
e^{i\delta_l}i^{-l} \
Y_{lm}(\n)\,
\left(\begin{array}{rrr}
l_f&1&l\\
m_f&M_P&m\\
\end{array}\right)
\la kl \, n_fl_f\| \ d \ \|\Psi_0\ra
$$
we can transform \Eref{TDCS} into the form:
\ba
\label{TDCS1}
\frac{d^3 \sigma_{2\gamma}}{d\Omega_1 d\Omega_2\,dE_2}&=&
C_{2\gamma}
\omega^\beta  \
\Bigg|
\sum_{J=0,2}
\sum_{l_1 l_2}
\{\e\otimes\e\}_J\cdot{\cal Y}^{\ l_1l_2}_J(\n_1\n_2)
\\ &&\hs{2cm}\times
(-i)^{l_1+l_2}
e^{i[\delta_{l_1}(k_1)+\delta_{l_2}(k_2)]} \
{\cal D}^{(2)}_{l_1l_2\,J}(k_1,k_2)
\Bigg|^2
\nonumber
\ea
Here we introduced  a bipolar  harmonic 
\cite{V88}
\be 
\hs{-2cm}
{\cal Y}_{JM}^{l_1l_2}(\n_1 , \n_2) =
\{
Y_{l_1}(\n_1)\otimes
Y_{l_2}(\n_2)
\}_{JM}=
\sum_{m_1m_2}
C^{JM}_{l_1m_2,l_2m_2} Y_{l_1m_1}(\n_1)Y_{l_2m_2}(\n_2)
\ee
The unit vectors $\n_i=\k_i/k_i$ are directed along the photoelectron
momenta. The tensor $\{\e\otimes\e\}_{JM_J}=\sum_{M_PM'_P}
C^{JM_J}_{1M_P,1M'_P}e_{M_P}e_{M'_P} $ represents the polarization of
light.
The reduced matrix element of the TPDI is given by the expression:
\be
\label{E8}
\hs{-2cm}
{\cal D}^{(2)}_{l_1l_2\,J}(k_1,k_2)=
\sum_{il}\!
\int \! p^2dp 
\frac
{\langle 
\Psi_{fl_1(J)}(k_1)\| \, d \, \|\Psi_{il(J=1)}(p)\rangle
\langle \Psi_{il(J=1)}(p)\|\, d \, \|\Psi_0\rangle}
{E_0+\omega-p^2/2-\varepsilon_i+i\delta}\\
\langle l_2k_2\!\parallel\! f\rangle~,
\ee
Here the CCC two-electron state \eref{E7}, stripped of its angular
dependence, is defined as
\be
\label{stripped}
\hs{-1cm}
\Psi_{fl (J)}(k)
\equiv
\| kl \, n_fl_f \ra
+\sum_{jL'}\isum k'^2dk'
 {\la kl \, n_fl_f\| T_J \| n_jl_j k'l' \ra \over 
E-k'^2/2-\varepsilon_j+i0} \ 
\| k'l'\, n_jl_j \ra
\ee
The bare dipole matrix element between the correlated He atom ground
state and the CCC channel state $\la kl \, n_il_i\| d \|\Psi_0\ra$ is
evaluated elsewhere \cite{KB98d}. By similar technique, the bare dipole
matrix element between two CCC channel states $\la k_1l_1 \, n_fl_f \| d
\| n_il_i \, pl \ra$ breaks down into one-electron radial integrals
and simple angular coefficients.

Expression \eref{stripped} contains the dipole matrix elements between
two continuum states. This matrix elements are ill-defined in the
length and velocity gauges of the electromagnetic interaction. To deal
with these integrals, we use the so-called Kramers-Henneberger form of
the Hamiltonian describing interaction of the atom and the
electromagnetic field.  The matrix elements of the electromagnetic
interaction operator in this representation can be written as
\cite{IK05b}:
\be
\label{Kramers}
\hs{-2cm}
\left\langle a, n+p\left|\hat H_{\rm int}^{\rm KH}\right|
b, n\right\rangle=
\left({\omega^2\over F}\right)^p
{1\over \pi} \sum\limits_{i=1}^2
\int\limits_0^{\pi}
\cos{p\theta}
\left\langle a\left|\ {Z\over r_i}-{Z\over |\bm r_i+
{\bm F}\cos{\theta}/ \omega^2|}\right|b\right\rangle
\ d\theta
\ee
We employ here the notation $|a, m\rangle$ where $a$ stands for a set
of quantum numbers describing the atom and $m$ denotes a number of
laser photons in a given mode. As can be seen, matrix elements defined
by the \Eref{Kramers} are finite and well-defined even if both $a$
and $b$ are continuum atomic states.  In Eq.(\ref{Kramers}) $\bm F$ is
the classical field strength. It is a vector directed along the
polarization vector of light, its magnitude is related to the photon
density as $F^2/8\pi = n\omega$. Both matrix elements in
\Eref{TDCS} correspond to the dipole transitions with $p=1$. 
For such a dipole transitions, in the weak field limit $F\to 0$,
operator \eref{Kramers} coincides, with the dipole operator in the
acceleration gauge $Z\r_i/r_i^3$.  In addition, operator
\eref{Kramers} can connect directly the initial and the final states
in \Eref{TDCS} by simultaneous absorption of two photons ($p=2$). Both
sequential and simultaneous absorption processes are of the same order
and should be included in the second order perturbation
theory. See
\cite{IK05b} for detail.

If we assume that the integrand in \Eref{E8} is a smooth function,
we can take out an average energy denominator and use the completeness 
of the CCC basis. This procedure, known as the closure approximation,
will take us to the following result:
\be
\label{closure}
{\cal D}^{(2)}_{l_1l_2\,J}(k_1,k_2)=
\Delta^{-1}
\langle 
\Psi_{l_1f(J)}(k_1)\| \ d \,  d \ \|\Psi_0\rangle
\langle l_2k_2\parallel f\rangle~,
\ee
The reduced matrix elements entering \Eref{closure}, in the length
gauge, have been worked out in an earlier paper on the second Born
treatment of the electron impact ionization-excitation and double
ionization of He
\cite{K04}.

By integrating TDCS \eref{TDCS1} over the angles $\Omega_1,\Omega_2$
one gets the single differential, with respect to the energy,
cross-section (SDCS). In the closure approximation, it is given by the
following expression:
\begin{equation}
\label{SDCS}
\hs{-1cm}
\frac{d \sigma_{2\gamma}}{dE_2}
=
C_{2\gamma}
{\omega^2  \over \Delta^2} \
\frac{1}{2\sqrt{E_2}}
\sum_{J=0,2}
\Big| C^{J0}_{10,10}\Big|^2
\sum_{l_1l_2}
\Big|
\Psi_{l_1f(J)}(k_1)\| \ d \,  d \ \|\Psi_0\rangle
\langle l_2k_2\parallel f\rangle
\Big|^2
\end{equation}
Implicit in the above is that $E_1+E_2=E_0+2\omega$ with
$E_2=\epsilon_{n_2l_2}$ for some pseudostate $n_2$ and every
$l_2$. Variation of the Laguerre exponential fall-off for each $l_2$
allows this for one value of $E_2$ ~\cite{BF95sdcs}.

Further integration of \Eref{SDCS} over the energy $E_2$ leads to the
total integrated TPDI cross-sections (TICS):
\be
\label{TICS}
\sigma_{2\gamma}(\omega)
=
C_{2\gamma}
{\omega^2  \over \Delta^2}
\sum_{J=0,2}
\Big| C^{J0}_{10,10}\Big|^2
\sum_{fL}
\Big|
\langle 
\Psi_{fL(J)}(k)\| \ d \,  d \ \|\Psi_0\rangle
\Big|^2
\ee
In this expression, the energy of the pseudostate and the Coulomb wave
are bound by the energy conservation
$\epsilon_{f}+k^2/2=2\omega+E_0$.  When evaluating expressions
\eref{SDCS} and \eref{TICS}, we can assume that the largest
contribution to the sum over the intermediate states in
\eref{E8} comes from  those terms in which the energy of 
the continuum electron in the intermediate state is of the order of
the energy of the ejected photoelectron $p^2/2+\epsilon_i\simeq
k_f^2/2+\epsilon_f=E_0+2\omega$. Therefore we can aprroximate the ratio
by $\Delta^2/\omega^2\simeq 1$.

\subsection{Two-photon double ionization amplitudes}

Although expression \eref{TDCS1} has been derived in the second-order
perturbation theory, its tensorial structure is general for the
TPDI. It allows to introduce a simple parametrization of the TDCS in
the manner suggested for the one-photon double ionization by Huetz and
co-workers \cite{HSWM91,MSH97,MSLMH97}. We proceed as follows. We
rewrite \Eref{TDCS1} as
\ba
\hs{-2cm}
\frac{d^3 \sigma_{2\gamma}}{d\Omega_1 d\Omega_2\,dE_1}&\propto&
\Bigg|
\sum_{l_1 l_2}
\b_0\cdot{\cal Y}^{\ l_1l_2}_0(\n_1\n_2)
 M_{l_1l_2}(k_1,k_2)
\\ && \hs{1cm}+
\b_2\cdot{\cal Y}^{\ l_1l_2}_2(\n_1\n_2)
Q_{l_1l_2}(k_1,k_2)
\Bigg|^2  
\equiv
  \Bigg|A_M+A_Q\Bigg|^2
\ea
Here we introduced the monopole $M_{l_1l_2}(k_1,k_2)$ and quadrupole
$Q_{l_1l_2}(k_1,k_2)$ reduced matrix elements modified by the
phase factors and overlaps. The tensor $\b_J=\{\e\otimes\e\}_J$
represents the polarization of light.

We first deal with the quadrupole amplitude
\ba
A_Q&=&
\Bigg\{
 \sum_{l_1 = l_2} 
+\sum_{l_1 = 0 \atop l_2=l_1+2}
+\sum_{l_2 = 0 \atop l_1=l_1+2}
\Bigg\}
\b_2\cdot {\cal Y}^{\ l_1l_2}_2(\n_1\n_2)
Q_{l_1l_2}(k_1,k_2)
\\&=&
\sum_{l = 0} 
\b_2\cdot{\cal Y}^{\ ll}_2(\n_1\n_2)
Q_{ll}(k_1,k_2)
\nonumber
\\&&\hs{0cm}+ 
\sum_{l = 0} 
\b_2\cdot
\Bigg\{
{\cal Y}^{\ ll+2}_2(\n_1\n_2)
Q_{ll+2}(k_1,k_2)
+
{\cal Y}^{\ l+2 \ l}_2(\n_1\n_2)
Q_{l+2 \ l}(k_1,k_2)
\Bigg\}
\nonumber
\ea
We introduce the symmetrized quadrupole matrix elements:
\be
Q^\pm_{l_1l_2}(k_1,k_2) = 
\frac12
\left\{
Q_{l_1l_2}(k_1,k_2)\pm
Q_{l_1l_2}(k_2,k_1)
\right\}
=\pm Q^\pm_{l_1l_2}(k_2,k_1)
\ee
Using this notation and the symmetry property ${\cal
Y}^{l+2l}_J(\n_1\n_2) = {\cal Y}^{ll+2}_J(\n_2\n_1)$, we can write the
quadrupole amplitude as
\ba
A_Q&=&
\sum_{l = 0} 
\b_2\cdot{\cal Y}^{\ ll}_2(\n_1\n_2)
Q_{ll}(k_1,k_2)
\\&+&
\sum_{l = 0} 
Q^+_{l_1l_2}(k_1,k_2)
\Big\{
\b_2\cdot{\cal Y}^{\ ll+2}_2(\n_1\n_2)
+
\b_2\cdot{\cal Y}^{\ ll+2}_2(\n_2\n_1)
\Big\}
\nonumber
\\&+&
\sum_{l = 0} 
Q^-_{l_1l_2}(k_1,k_2)
\Big\{
\b_2\cdot{\cal Y}^{\ ll+2}_2(\n_1\n_2)
-
\b_2\cdot{\cal Y}^{\ ll+2}_2(\n_2\n_1)
\Big\}
\nonumber
\ea
We use the bipolar harmonics expressions given by \citeasnoun{MMM96}:
\ba
\label{ll}
\hs{-2cm}
{\cal Y}^{ll}_2(\n_1,\n_2) &=&
C_l
\Bigg\{
P'_l(x)
\{\n_1\otimes\n_2\}_2+
P''_l(x)
\{[\n_1\times\n_2]\otimes[\n_1\times\n_2]\}_2
\Big\}
\\
\label{ll+2}
\hs{-2cm}
{\cal Y}^{ll+2}_2(\n_1,\n_2) &=&
D_l
\Bigg\{
  P''_l(x) \{\n_1\otimes\n_1\}_2
+ P''_{l+2}(x) \{\n_2\otimes\n_2\}_2
\\&&\hs{4cm}
-2P''_{l+1}(x) \{\n_1\otimes\n_2\}_2
\Big\}
\nonumber
\ea
where $x=\cos\theta_{12}=\n_1\cdot\n_2$ and $C_l$, $D_l$ are the
normalization coefficients.  These expressions allow for the following
parametrization of the quadrupole amplitude:
\ba
\hs{-2cm}
\label{parametrization}
A_Q
&\equiv&
g^+ 
\Big[\{\n_1\otimes\n_1\}_2+\{\n_2\otimes\n_2\}_2\Big]\cdot\b_2
+
g^- 
\Big[\{\n_1\otimes\n_1\}_2-\{\n_2\otimes\n_2\}_2\Big]\cdot\b_2
\\
\hs{-2cm}
&&\hs{2cm}+
g_s \{\n_1\otimes\n_2\}_2\cdot\b_2
\nonumber
+
\nonumber
g_0
\{[\n_1\times\n_2]\otimes[\n_1\times\n_2]\}_2\cdot\b_2
\ea
Here we introduced the four symmetrized amplitudes:
\ba
\label{amplitudes}
g^\pm(k_1,k_2,x)&=&
\sum_{l=0}
D_l 
\Big[P''_{l}(x) + P''_{l+2}(x)\Big]
Q^\pm_{ll+2}(k_1,k_2)
\nonumber
\\
g_s(k_1,k_2,x)&=&
\sum_{l=1}
C_l P'_{l}(x) Q_{ll}(k_1,k_2)
-
4D_l P''_{l+1}(x) Q^+_{ll+2}(k_1,k_2)
\\
g_0(k_1,k_2,x)&=&
\sum_{l=2}
C_l P''_{l}(x) Q_{ll}(k_1,k_2)
\nonumber
\ea
For equal energy sharing $g^-(k_1=k_2)=0$. \Eref{parametrization}
allows to separate the kinematic variables represented by the tensor
products of vectors $\n_1$ and $\n_2$ and the dynamics of
two-electron escape which is contained in the TPDI amplitudes.

For the coplanar geometry, when the two photoelectron momenta and the
photon polarization vector belong to the same plane, we can make the
following choice of the coordinate frame: $z\ \| \ \k_1$ and
$y \ \| \ [\k_1\times\k_2] $. In this case the quadrupole amplitude is
given by the formula
\ba
\label{parametrization1}
A_Q&=&
\frac23
\Bigg\{
g^+
\Big[P_2(\cos\theta_1)+P_2(\cos\theta_2)\Big]
+
g^-
\Big[P_2(\cos\theta_1)-P_2(\cos\theta_2)\Big]
\\&&\hs{3cm}+
\frac12
g_s
\Big[3\cos\theta_1\cos\theta_2 - x\Big]
+
\frac12
g_0
(x^2-1)
\Bigg\}
\nonumber
\ea
where $\cos\theta_i = \n_i\cdot \e$, $x=\cos(\theta_2-\theta_1)$.
The monopole amplitude  is given by the expression:
\be
\hs{-2.5cm}
f_0=
\sum_{l_1= l_2}
\b_0\cdot{\cal Y}^{\ l_1l_2}_0(\n_1\n_2)
M_{l_1l_2}(k_1,k_2) 
=
-{1\over\sqrt3}
\sum_{l = 0} 
(-1)^l{\sqrt{2l+1}\over4\pi}
P_l(x) \  M_{ll}(k_1,k_2)
\ee
So, in case of an arbitrary energy sharing, the theoretical
description of the TPDI TDCS requires five amplitudes as compared with
only two amplitudes needed to describe the one-photon double
ionization TDCS.  This reflects a much richer dynamical structure of
the two-electron continuum

\section{Numerical results}
\label{Sec3}

\subsection{Integrated cross-sections}
\label{Sec3a}

\begin{figure}[h]
\epsfxsize=5.5cm
\epsffile{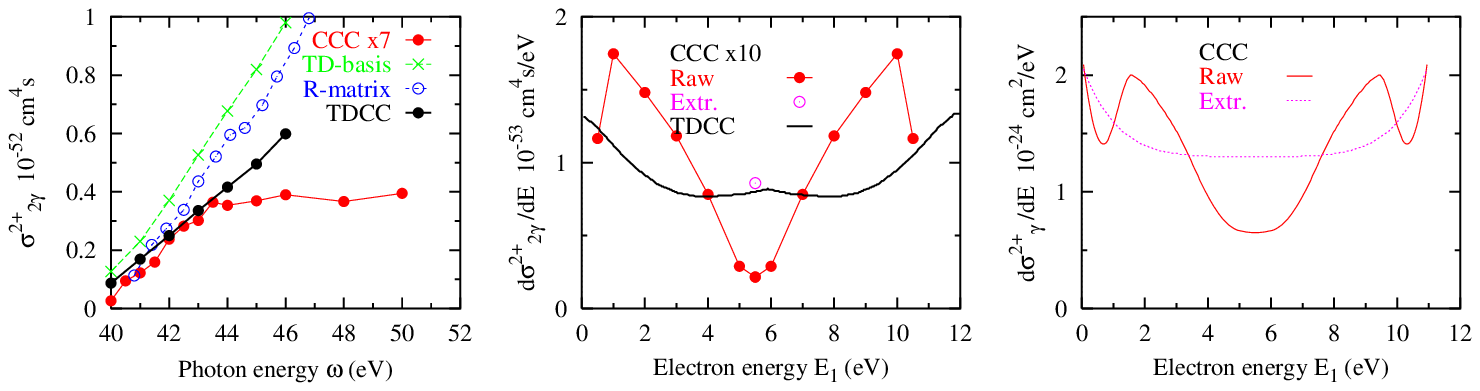}
\caption{
\label{Fig1}
{\em Left panel:}
Total cross-section of the TPDI on He as a function of the photon
energy~ $\omega$. Present CCC closure calculation is compared with
the TD-calculation of \citeasnoun{PBLB03}, $R$-matrix calculation of
\citeasnoun{FdH03} and Grid calculation of \citeasnoun{HCC05}. 
{\em Central panel:} Single differential cross-section of the TPDI of
He at $\omega=45$~eV. The extrapolated CCC SDCS at equal energy
sharing is shown by an open circle. The TDCC results are from
\citeasnoun{CP02a}. {\em Right panel:} Single differential cross-section
of the single-photon double ionization of He at the same excess
energy. A ``raw'' and extrapolated CCC calculations are shown by the
red solid and purple dashed lines, respectively.}
\end{figure}

We start presenting our numerical results with the total integrated
cross-section (TICS) of the TPDI of He.  On the left panel of
\Fref{Fig1} we show the TPDI TICS from several  closure 
calculations at various photon energies in comparison with the
literature values. As is seen from the figure, 
the presently calculated cross-section falls substantially below
predictions of other methods. 
We cannot say whether this is a shortcoming of the closure
approximation or the second-order perturbation theory in general. The
full CCC calculation of the total cross-section is too much time
consuming to get the data across a wide photon energy range.

Evaluation of the magnitude of the TDCS requires the knowledge of the
single-differential cross-section (SDCS) resolved with respect to the
photoelectron energy. As was shown in the case of the single-photon
double ionization, the ``raw'' SDCS, as calculated by the CCC method,
is oscillatory due to explicit distinguishability of the two
photoelectrons. A special extrapolation procedure was devised to cure
these unphysical oscillations which relied on the fact that the area
beneath the SDCS is the total cross-section and the  equal energy
SDCS should be given by the coherent sum of the direct and exchange
CCC amplitudes \cite{KB02}. The raw TPDI SDCS from the CCC
closure calculation at $\omega=45$~eV is shown in the central panel of 
\Fref{Fig1} in comparison with the TDCC calculation of
\citeasnoun{HCC05}. The full extrapolation procedure was not
implemented in the case of TPDI except for the midpoint which was
calculated with the coherent sum of the direct and exchange
amplitudes. 
For comparison, in the right panel of \Fref{Fig1} we show
the SDCS of the single-photon double ionization of He at the same
excess energy. Both the raw and extrapolated calculations are shown
which resemble respectively the CCC and TDCC TPDI results on the
central panel of \Fref{Fig1}.

\subsection{Triple differential cross-sections}
\label{Sec3b}

\begin{figure}[h]
\epsfxsize=4.5cm
\epsffile{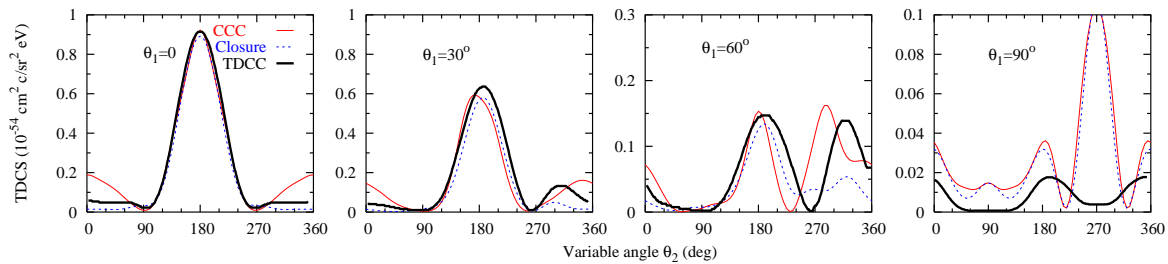}
\caption{
\label{Fig2}
TDCS TPDI of He for the coplanar geometry at $\omega=45$~eV and
$E_1=E_2=5.5$~eV.  A $17l4$ full CCC calculation (red solid line) and
a $20l5$ closure calculation (blue dashed line) are compared with the
TDCC calculation of \citeasnoun{CP02a} (black thick solid line). }
\end{figure}

The full CCC calculation of the TDCS TPDI based on Eqs.~\eref{TDCS1}
and \eref{E8} is very time consuming. So far, we were able to perform
just one such calculation at the photon energy of 45~eV and equal
energy sharing of the two photoelectrons $E_1=E_2=5.5$~eV. The
resulting TDCS are shown in \Fref{Fig2} in comparison with the TDCC
calculation of \citeasnoun{CP02a}. Here the coplanar geometry is
assumed with one electron escaping at a fixed angle $\theta_1$ and the
second electron detected on the full angular range. The CCC basis of
the final state included $17-l$ target states with orbital momentum
$l$ ranging from 0 to 4 (the so-called $17l4$ calculation). The CCC
basis of the intermediate states was much shorter with only three
target states ($1s,2s,2p$) included. This was possible because the
absorption of a single 45~eV photon could populate only few target
states.

As it will be shown in the subsequent analysis, the TDCS is dominated
by the $D$-wave except for the special case of the fixed electron
angle $\theta_1=90^\circ$. In this case a very small TDCS is a result
of a delicate compensation of the $D$ and $S$-waves. The CCC
calculation reproduces the TDCC calculation of \citeasnoun{CP02a}
fairly well except for this special case when the $D$ wave remains
uncompensated and produces a very large peak.

The closure approximation \eref{closure} makes a considerable
simplification of the calculation and a much larger basis of the final
target states can be handled with the present computational resources.
In \Fref{Fig2} we show the results of a $20l5$ closure calculation at
the same photon energy $\omega=45$~eV and $E_1=E_2=5.5$~eV. 
At $\theta_1=0^\circ$, both the full CCC calculation and the closure
calculation perform fairly well, the closure calculation resulting in
a somewhat lesser ``wings'' on the sides of the main peak. At
$\theta_1=30^\circ$ and especially at $\theta_1=60^\circ$, the full
CCC calculation reproduces better a second peak appearing at around
$\theta_2\simeq270^\circ$. In the most difficult case of
$\theta_1=90^\circ$, both calculations produce a large spurious peak at
$\theta_2\simeq270^\circ$ which is not visible in the TDCC
calculation. As will be discussed in the following, this peak is a
result of an undercompensated $D$ wave.  Generally, the closure
approximation results are quite close to the full CCC calculation. In
the rest of the paper, we will focus on the closure approximation
which is much less computationally demanding.

\begin{figure}[h]
\vspace*{6.5cm}
\epsfxsize=4.5cm
\epsffile{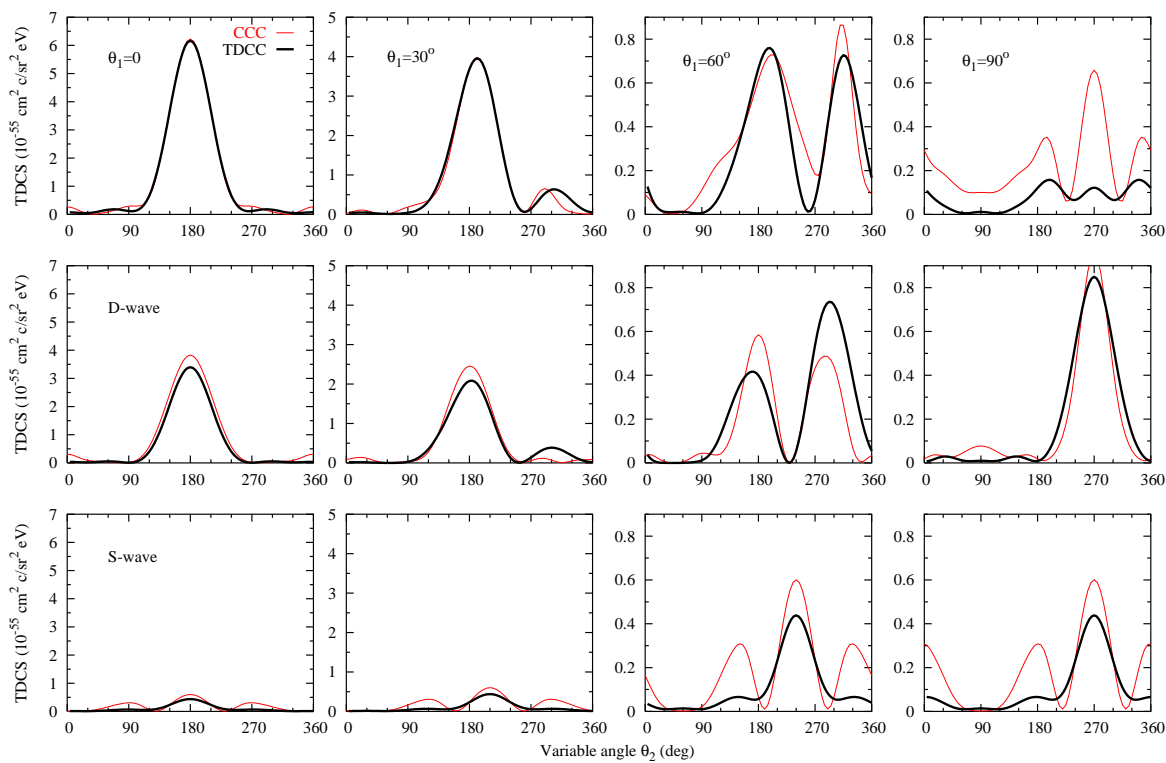}
\caption{
\label{Fig3}
TDCS TPDI of He for the coplanar geometry at $\omega=42$~eV and
$E_1=E_2=2.5$~eV.  A $20l6$ closure calculation (red solid line) is
compared with the TDCC calculation of \citeasnoun{HCC05} (black thick
solid line). The top row shows the TDCS with combined contributions of 
$D$- and $S$-waves whereas in the middle and bottom rows the separate
contribution of the $D$ and $S$-waves are plotted. }
\end{figure}

Further analysis of the TPDI TDCS is carried out in \Fref{Fig3} where
we present a set of TDCS for the photon energy $\omega=42$~eV and
$E_1=E_2=2.5$~eV. Following \citeasnoun{HCC05}, we give a separate
contribution of the $D$-and $S$-waves to the TDCS at various fixed
angles $\theta_1$.  As is seen in the figure, the $D$-wave
contribution is dominant for all fixed electron angles except for
$\theta_1=90^\circ$. Here a very small cross-section of
\citeasnoun{HCC05} is result of an almost complete compensation of the
partial $S$- and $D$-wave contributions.  The present CCC closure
calculation deviate insignificantly from
\citeasnoun{HCC05} but no such a perfect compensation of the partial
waves occurs. As the result, the TDCS at  $\theta_1=90^\circ$ is quite 
different from the TDCC calculation.

\begin{figure}[h]
\epsfxsize=4.5cm
\epsffile{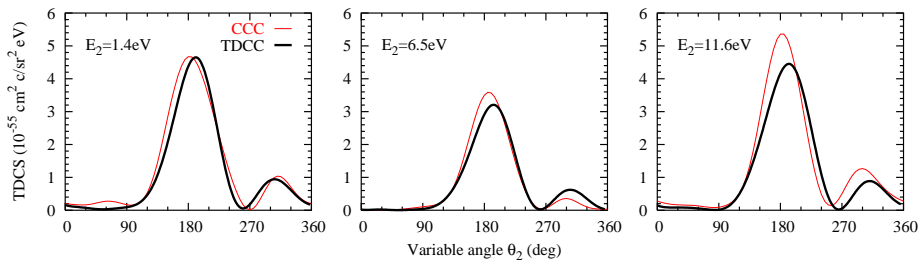}
\caption{
\label{Fig4}
TDCS TPDI of He for the coplanar geometry at $\omega=46$~eV and
various energy sharings. A $20l5$ closure calculation (red solid line)
is compared with the TDCC calculation of \citeasnoun{HCC05} (black
thick solid line). In all panels, the fixed electron angle is
$\theta_1=30^\circ$.}
\end{figure}

More data is presented in \Fref{Fig4} where we show the TDCS at the
photon energy $\omega=46$~eV and various, both equal and unequal,
energy sharings. Here the fixed electron angle is kept constant at 
$\theta_1=30^\circ$. Generally, a good agreement with the TDCC
calculation of  \citeasnoun{HCC05} is seen with some minor shape
variations. This has been the case previously as at this particular
value of the fixed electron angle the main contribution to the TDCS
comes from the $D$-wave and no significant interference between the
partial waves occurs.

As it was pointed out when discussing the SDCS calculations, the equal
energy sharing TDCS can be calculated fully {\em ab initio} in the CCC
method by constructing the coherent sum of the direct and exchange
ionization amplitudes. Therefore, all the TDCS shown in
Figs.~\ref{Fig2}-\ref{Fig3} are absolute. When making comparison with
the TDCC results, only one scaling constant
was used across all the panels of these figures to account for
somewhat different total integrated cross-sections in the two methods.
For the asymmetric energy sharing, however, the magnitude of the CCC
TDCS can be affected by the SDCS oscillations. Therefore, the CCC
results in
\Fref{Fig4} were scaled individually on the side and central panels to
make a shape comparison with the TDCC data of
\citeasnoun{HCC05}.

\subsection{TPDI amplitudes}
\label{Sec3c}

\begin{figure}[h]
\epsfxsize=4.5cm
\epsffile{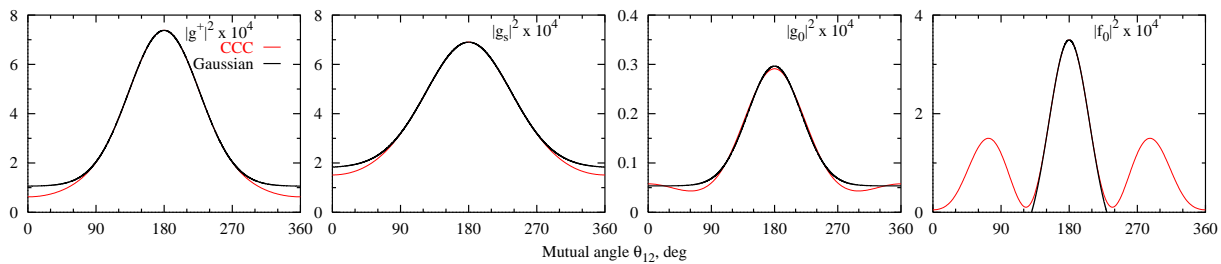}

\caption{
\label{Fig5}
Symmetrized amplitudes $g^+$, $g_s$, $g_0$  and $f_0$ as functions of
the mutual electron angle $\theta_{12}$  for
$\omega=45$~eV and $E_1=E_2=5.5$~eV.  
}
\end{figure}

The TDCS presented in the preceding section can be most conveniently
analyzed using parametrization \eref{parametrization1} with the
symmetrized amplitudes defined by \Eref{amplitudes}. In the case of
equal energy sharing, only three quadrupole amplitudes and one
monopole amplitude are needed. The moduli squared of these amplitudes
are shown in \Fref{Fig5} for $\omega=45$~eV and $E_1=E_2=5.5$~eV.

As in the case of the single-photon double ionization, all the
amplitudes peaks strongly at $\theta_{12}=180^\circ$.  This is a
result of the strong electron repulsion favoring back-to-back
emission. The central part of the amplitudes around
$\theta_{12}=180^\circ$ was fitted with the Gaussian  on a flat
pedestal:
\be
\label{Gaussian}
 |g|^2  \propto
a\exp\left[ -4\ln2{(\pi-\theta_{12})^2\over\Delta\theta_{12}^2}
\right] + b
\ee
The fitting seems to be nearly perfect with the width parameters
$\Delta\theta_{12}$ of $110^\circ$, $131^\circ$, $63^\circ$ and
$105^\circ$ for $g^+$, $g_s$, $g_0$ and $f_0$, respectively. For
comparison, at the present energy sharing, the symmetric amplitude of
the single-photon double ionization has the Gaussian width of
$96^\circ$, somewhat less than  that of $g^+$.

 Gaussian ansatz \eref{Gaussian} combined with
\Eref{parametrization1} allows to analyze the evolution of the TPDI
TDCS as a function of the fixed photoelectron angle. At $\theta_1=0$,
$\theta_{12}=\theta_2$ and
$$
P_2(\cos\theta_1)+P_2(\cos\theta_2) = \frac12
(3\cos\theta^2+1) \ \ , \ \ 
 \frac12(3\cos\theta_1\cos\theta_2 - x) = 
\cos\theta_2 
$$
We see that the kinematic factors accompanying the largest amplitudes
$g^+$ and $g_s$ both peak at $\theta_2=180^\circ$ where the amplitudes
have the maximum. This produces a bold peak seen in all TDCS figures
at this combination of angles $\theta_1$ and $\theta_2$. This is in
sharp contrast to one-photon double ionization in which the kinematic
factor is represented by $P_1(\cos\theta_1)+P_1(\cos\theta_2)$ which
has a node at $\theta_{12}= 180^\circ$. As a result, the
one-photon TDCS has a maximum at a compromise angle where neither the
kinematic factor nor the amplitude have their respective maxima.

The shape of the TPDI TDCS changes completely at $\theta_1=90^\circ$
where
$$
P_2(\cos\theta_1)+P_2(\cos\theta_2) = \frac12
(3\cos^2\theta_2-2) \ \ , \ \ 
 \frac12(3\cos\theta_1\cos\theta_2 - x) = 
-\frac12\sin\theta_2 
$$
Neither of the factors peak at $180^\circ$ and the $D$-wave
contribution has the maximum at a compromise angle of $270^\circ$.
A partial overlap between the kinematic and dynamic factors
result in a much smaller cross-section.  Coincidentally, the $S$-wave
has the peak at $\theta_2=270^\circ$ as well. However, the relative
phase of the $S$ and $D$ amplitudes conspires to cancel out the two
contributions. This cancellation is a dynamic property depending on
the relative magnitude of the quadrupole and monopole amplitudes.
It is somewhat different in the CCC and TDCC calculations which assume 
different models of the two-electron dynamics.

\begin{figure}[h]
\epsfxsize=4.5cm
\epsffile{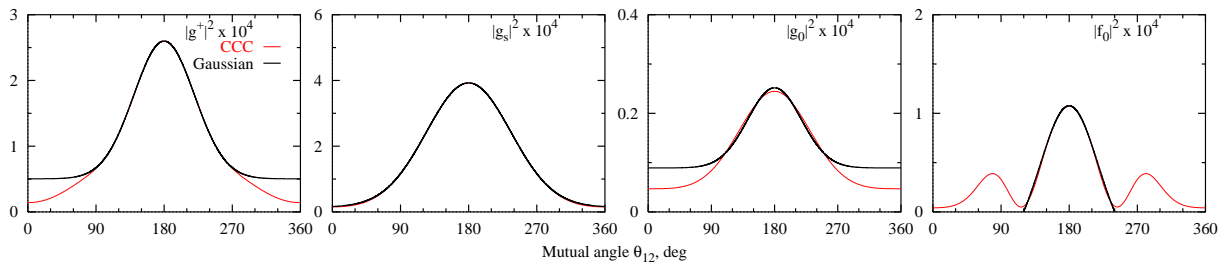}

\caption{
\label{Fig6}
Symmetrized amplitudes $g^+$, $g_s$, $g_0$ and $f_0$ as functions of
the mutual electron angle $\theta_{12}$ for $E_1=E_2=20$~eV.  }
\end{figure}

In \Fref{Fig6} we analyze the TPDI amplitudes much further away from
the double ionization threshold at the equal energy sharing of
$E_1=E_2=20$~eV.  All the amplitudes were fitted with the Gaussian
ansatz.  The width parameters are $94^\circ$, $131^\circ$, $90^\circ$
and $90^\circ$ for $g^+$, $g_s$, $g_0$ and $f_0$ respectively.  There
is no systematic change of the Gaussian width parameters at 40~eV as
compared to 4~eV excess energy. In the meantime, the Gaussian width of
the single-photon ionization amplitude has increased from $96^\circ$
to $103^\circ$. 

Qualitative similarity of the amplitudes in \Fref{Fig5} and
\Fref{Fig6} allows us to predict that the basic shape of the TDCS will
remain unchanged much further away from the double ionization
threshold compared to what was investigated by \citeasnoun{HCC05}.  To
make more quantitative predictions for the TDCS a more systematic
study of the amplitudes is needed including a thorough test of
convergence with respect to the size of the CCC basis.

\vspace*{-1cm}
\section{Conclusion}
\label{Sec4}

In the present paper we investigated the two-photon double ionization
of He in a wide range of photon energies from 4 to 40~eV above the
threshold. The electron-photon interaction was treated to the lowest
second order whereas the electron-electron interaction was included
non-perturbatively.  We applied the convergent close-coupling
formalism to describe the $S$- and $D$ two-electron continua of the
final doubly ionized state. The intermediate $P$-state was also
described with the CCC expansion, albeit of a smaller size. The
ill-defined continuum-continuum dipole matrix elements between the
intermediate and final states were handled using the
Kramers-Henneberger gauge of the electromagnetic interaction.  As an
alternative and much less computationally extensive solution, we used
the closure approximation summing over the complete set of the
intermediate states. By doing so, we were able to reduce the TPDI
amplitude to the matrix element of the squared dipole operator, in the
length gauge, between the correlated ground state and the two-electron
CCC final state.

Near the threshold, the presently calculated total integrated TPDI
cross-section is of the same order of magnitude as predicted by other
non-perturbative methods. However, further away from the threshold,
the CCC closure results fall substantially below the reported
literature values. The single differential, with respect to the
energy, cross-section was also evaluated. As in the single-photon
double ionization, the raw CCC SDCS shows unphysical oscillations and
needs rescaling. However, the equal energy sharing point can still be
calculated {\em ab initio}. 

The fully-resolved triple differential cross-sections were evaluated
using integration over the complete set of CCC intermediate state as
well as the closure approximation. Both sets of results agree
reasonably well with the previously  reported TDCC cross-sections. 
The only exception is the fixed electron angle of 90$^\circ$ where the 
delicate cancellation of the $S$- and $D$-partial waves takes place.
Here the CCC and TDCSS results are at variance. These findings
indicate that the angular correlation pattern in TPDI is formed mainly 
due to the inter-electron interaction in the two-electron
continua. Influence of the strong laser field and precise mechanisms of 
the two-photon absorption is less important. Of course, this
conclusion needs to be further confirmed experimentally.

A complete set of symmetrized amplitudes was introduced to parametrize
conveniently the TPDI TDCS. The amplitudes display a Gaussian shape
and allow to explain the evolution of the TDCS at varying fixed
electron angles. The amplitudes evaluated at fairly high excess energy
of 40~eV resemble closely the amplitudes near threshold. This allows
to predict a rather robust shape of the TPDI TDCS across a wide range
of photon energies. Reported Gaussian parameters can be useful in
modeling the TPDI TDCS at various experimental kinematics.  A more
systematic and consistent study is required, and is planned, to serve
this purpose.  Numerical values reported here serve only as an
illustration and should be treated as preliminary because not all the
convergence checks were possible to perform at the present
stage. Therefore, some minor variation in the width parameters might
occur as the size of the CCC basis varies.

This work is the first report on the application of the CCC method to
the TPDI process in He. We intend to continue this work, both within
the scope of the perturbation theory and treating the electromagnetic
field non-perturbatively.  The basic framework of the theory is
outlined in our earlier paper \cite{IK05}.  The theory is potentially
capable of studying a wide range of photon energies and
electromagnetic field intensities.  The only limitation is available
computational resources which we hope to lift in the near future.

\section{Acknowledgements}

The authors wish to thank James Colgan for providing TDCC results in
numerical form.  The authors acknowledge support of the Australian
Research Council in the form of Discovery grant DP0451211. Facilities
of the Australian Partnership for Advanced Computing (APAC) were used.

\newpage

\end{document}